\documentclass[aps,prb,twocolumn,floatfix,superscriptaddress,showpacs]{revtex4}

\usepackage{graphicx}
\usepackage{dcolumn}
\usepackage{bm}
\usepackage{textcomp}
\usepackage{amssymb}

\begin{document}

\title{Scaling of the Quantum-Hall plateau-plateau transition in graphene}

\author{A.~J.~M.~Giesbers}
\affiliation{
High Field Magnet Laboratory, Institute for Molecules and Materials,
Radboud University Nijmegen, Toernooiveld 7, 6525 ED Nijmegen, The Netherlands
}

\author{U.~Zeitler}
\email[]{U.Zeitler@science.ru.nl}
\affiliation{
High Field Magnet Laboratory, Institute for Molecules and Materials,
Radboud University Nijmegen, Toernooiveld 7, 6525 ED Nijmegen, The Netherlands
}

\author{L.~A.~Ponomarenko}
\affiliation{
Department of Physics and Astronomy, University of Manchester, M13 9PL, Manchester, UK
}

\author{R.~Yang}
\affiliation{
Department of Physics and Astronomy, University of Manchester, M13 9PL, Manchester, UK
}

\author{K.~S.~Novoselov}
\affiliation{
Department of Physics and Astronomy, University of Manchester, M13 9PL, Manchester, UK
}

\author{A.~K.~Geim}
\affiliation{
Department of Physics and Astronomy, University of Manchester, M13 9PL, Manchester, UK
}

\author{J.~C.~Maan}
\affiliation{
High Field Magnet Laboratory, Institute for Molecules and Materials,
Radboud University Nijmegen, Toernooiveld 7, 6525 ED Nijmegen, The Netherlands
}

\date{\today}     


\begin{abstract}
The temperature dependence of the magneto-conductivity in graphene
shows that the widths of the longitudinal conductivity peaks, for the 
$N=1$ Landau level of electrons and holes, display a 
power-law behavior following $\Delta \nu \propto T^{\kappa}$ with a scaling 
exponent $\kappa = 0.37\pm0.05$. 
Similarly the maximum derivative of the quantum Hall plateau transitions $(d\sigma_{xy}/d\nu)^{max}$ scales as $T^{-\kappa}$ with a scaling 
exponent $\kappa = 0.41\pm0.04$ for both the first and second
electron and hole Landau level. These results 
confirm the universality of a critical scaling exponent. In the zeroth Landau 
level, however, the width and derivative are essentially temperature independent, which we 
explain by a temperature independent intrinsic length that obscures the 
expected universal scaling behavior of the zeroth Landau level. 
\end{abstract}


\maketitle

The integer quantum Hall effect in two-dimensional systems (2DES) is 
a direct consequence of the density of states of the Landau levels.
Electronic states in the tails of individual Landau levels
are localized and give rise to the quantized plateaus in the
Hall resistance. The states in the center of the Landau levels
are extended, because their wave functions are delocalized.
The delocalization in 2DESs is shown to be governed by a 
localization length, which 
decays exponentially away from the Landau level centers.\cite{wei, pruisken}
The critical scaling exponent, is found to be 
universal for traditional 2DESs.\cite{Huckestein} 

Recently a new type of 2DES, graphene, joined the playground of the 
quantum Hall physics. Its 
unconventional Landau level spectrum of massless chiral Dirac fermions 
leads to a new half-integer quantum Hall effect,\cite{NovoselovNature, ZhangNature}
which remains visible up to room temperature.\cite{RTNovoselov}
 
Here we investigate the scaling behavior of the quantum Hall 
plateau-plateau transitions in graphene. When changing the carrier 
concentration $n$ at a constant field, the peak width of the 
longitudinal conductivity, $\Delta \sigma_{xx}$, as well as the inverse 
slope of the Hall conductivity $d\nu/d\sigma_{xy}$ scale as $T^{\kappa}$. 
Our experimentally measured scaling exponent $\kappa=0.40 \pm 0.02$ is consistent 
with universal scaling theory.\cite{wei, pruisken, Huckestein, Koch2, Hohls2, Romer} 
The transition through the zeroth Landau level, however, shows no clear 
scaling behavior which we explain by a temperature independent intrinsic 
length scale governing scaling. 

The single layer graphene sample was made by micromechanical 
exfoliation of a crystal of natural graphite and subsequently 
contacted by gold contacts and patterned into a 1 $\mu$m wide Hall-bar 
by electron-beam lithography and reactive plasma etching.\cite{NovoselovScience} The 
structure was deposited on a 300 nm Si/SiO$_{2}$ substrate 
thereby forming a graphene ambipolar field effect transistor. 
Prior to the measurements the sample was annealed at 400 K, placing its 
charge neutrality point (CNP) at zero gate voltage with a mobility of 
$\mu=1.0$~m$^2$(Vs)$^{-1}$.

\begin{figure}[tb]  
  \begin{center}
  \includegraphics[width=0.9\linewidth]{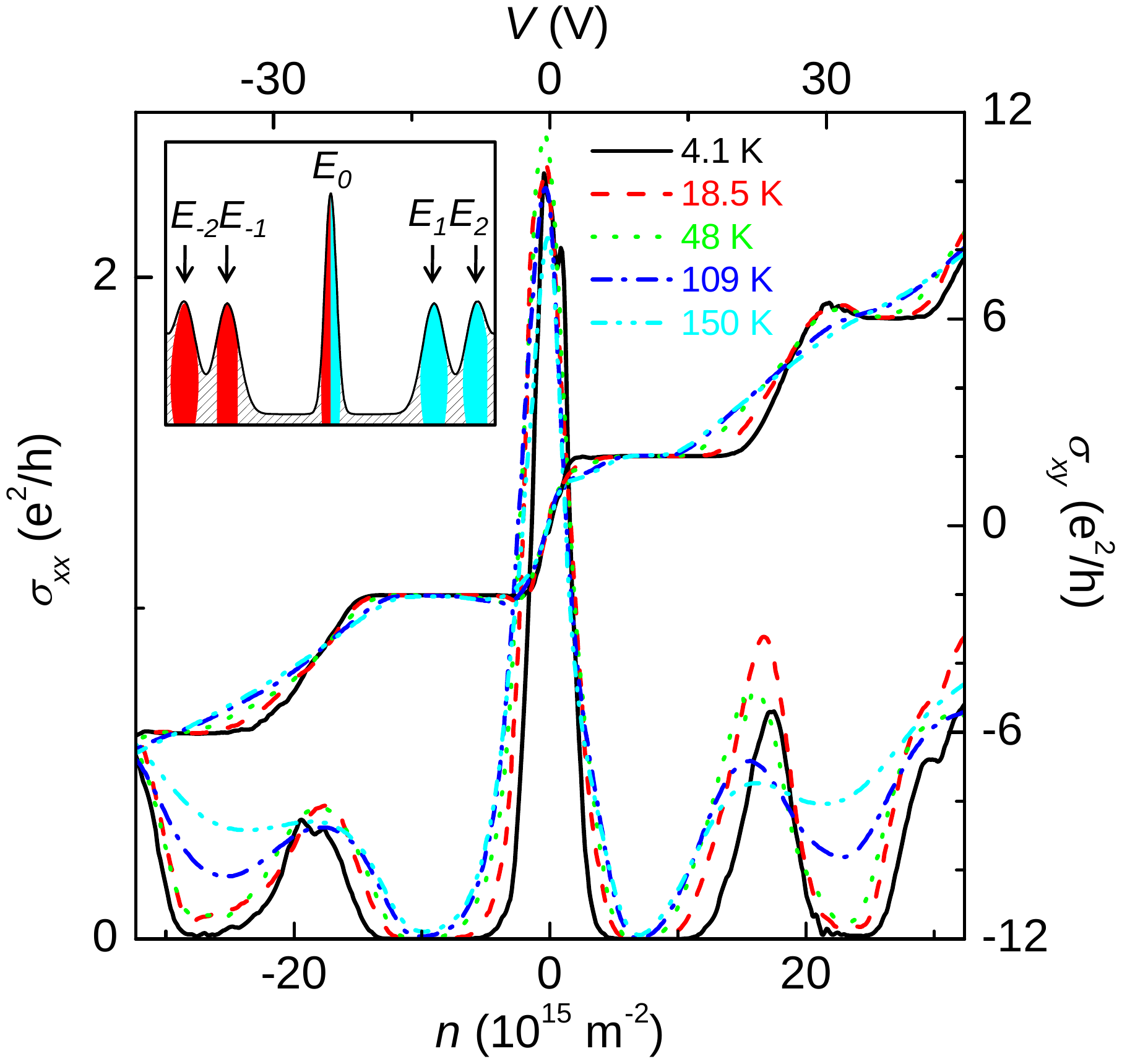}
  \end{center}
  \caption{(Color online) Longitudinal conductivity $\sigma_{xx}$ and 
  Hall conductivity $\sigma_{xy}$ as a function of concentration $n$ 
  (bottom-axis) and voltage $V$ (top-axis) at $B=20$~T for different 
  temperatures. The inset shows the Landau level spectrum in graphene.
  } 
  \label{Figure1}
\end{figure} 

In a magnetic field the density of states in graphene splits up into 
non-equidistant Landau levels\cite{NovoselovNature, ZhangNature, Gusynin, CNeto} 
according to $E_{n}=sgn(N)\sqrt{2 e \hbar v_{F}^{2} B |N|}$
(see inset Fig.~\ref{Figure1}), where $N$ identifies the 4-fold degenerate
Landau levels and is given by $N=0,\; \pm 1, \; \pm 2, \; \dots$. 
This Landau level spectrum leads to the half-integer quantum Hall 
effect as presented in Fig.~\ref{Figure1}, where we show the Hall conductivity
$\sigma_{xy}$ and longitudinal conductivity $\sigma_{xx}$ as calculated from 
the symmetrized resistivities by a tensor inversion. The 
electronic states between two Landau levels are localized (dashed regions 
in the inset of Fig.~\ref{Figure1}) 
and lead to Hall plateaus quantized to $\sigma_{xy}=4ie^{2}/h$, 
with $i=N+1/2$ a half integer. The plateaus are accompanied by zero minima in the 
longitudinal conductivity $\sigma_{xx}$. Around the centers of the Landau levels
the states are extended and the Hall conductivity changes to 
its next plateau, while the longitudinal conductivity displays a 
peak.\cite{note}

The region of extended states or delocalized states can be described by the
energy interval where the localization length $\xi(E)$, i.e., the spatial
extention of the wave function, increases beyond some characteristic 
length, $\xi(E) > L$. States remain localized if 
their localization length remains below this characteristic length, $\xi(E) < L$. 
According to scaling theory the localization length $\xi(E)$ follows a power law
behavior\cite{wei, pruisken} as a function of energy given by
\begin{equation}\label{loclength}
\xi(E)=\xi_{0}|E-E_{c}|^{-\gamma},
\end{equation}
with $E_{c}$ the singular energy at the center of the Landau level and 
$\gamma$ the critical exponent. Towards this singular energy the localization
length diverges and states become delocalized when $\xi(E) > L$, making the 
states in the center of the Landau levels extended.
To probe this delocalization we have to translate the energy relation to 
measurable quantities. In a first approximation, we assume a constant 
density of states (DOS) close to the singular energy.\cite{Hohls} 
In this case the energy in equation (\ref{loclength}) is directly 
proportional to the measurable filling factor, $\nu=nh/eB$, of the Landau 
levels, i.e. $E \propto \nu$. 

At finite temperatures the characteristic length $L$ is 
determined by the inelastic scattering rate, $L_{in}\propto T^{-p/2}$, with
$p$ the inelastic scattering exponent.
We can directly relate this temperature dependent quantity to the 
scaling of the localization length, and equation (\ref{loclength}) 
becomes a function of temperature, $|\nu-\nu_{c}| \propto T^{p/2\gamma}$, with
$\nu_{c}$ the Landau level center and $\nu$ its edge. 
This scaling behavior has therefore a direct relation to 
the quantum Hall transitions measured in the magneto-conductivity at 
finite temperatures through the quantities $(d\sigma_{xy}/d\nu)^{max}$ and 
$\Delta \nu$.\cite{wei, pruisken} The width $\Delta \nu$ is defined as the 
distance between two extrema in the derivative $d\sigma_{xx}/d\nu$ and 
$(d\sigma_{xy}/d\nu)^{max}$ is the maximum in the derivative of $\sigma_{xy}$. 
Both quantities show a power law behavior $\Delta \nu \propto T^{\kappa}$ and
$(d\sigma_{xy}/d\nu)^{max} \propto T^{-\kappa}$ with the critical 
exponents $\kappa$ and $\gamma$ related according to $\kappa=p/2\gamma$.
An exact value for $p$ can not be 
determined from our data, however $p=2$ is a commonly used value for
two-dimensional systems governed by short range scattering,\cite{Brandes, Koch1, wei1} 
which is indeed expected in our high mobility graphene 
sample.\cite{Tan, Chen, Peres1, Trushin} 

\begin{figure}[tb]  
  \begin{center}
  \includegraphics[width=0.9\linewidth]{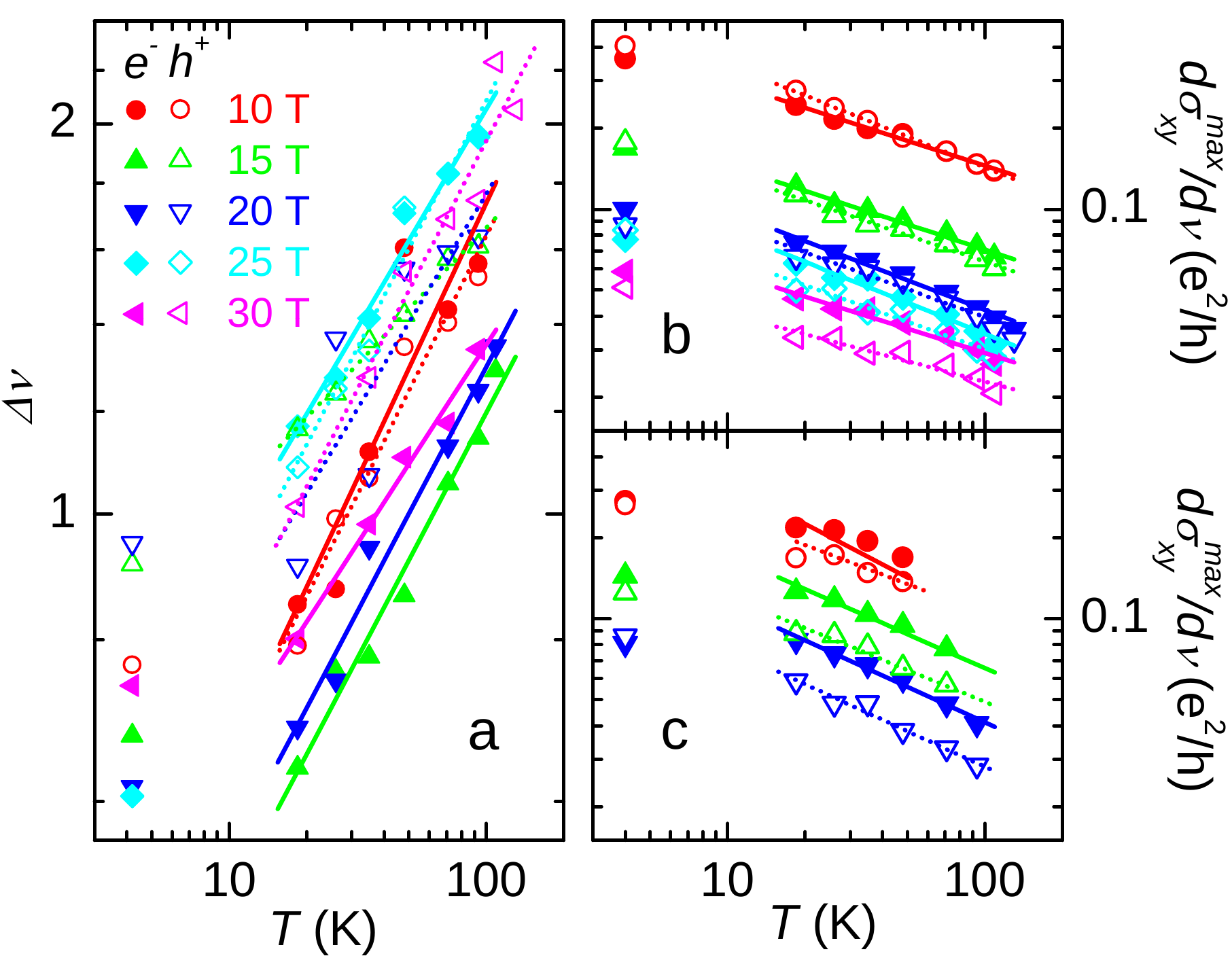}
  \end{center}
  \caption{(Color online) (a) $\Delta \nu$, distance between two extrema 
  in the derivative $d\sigma_{xx}/d\nu$, of the first electron Landau 
  level (solid symbols) and the first hole (open symbols) Landau level 
  for different magnetic fields (b) The derivative $(d\sigma_{xy}/d\nu)^{max}$
  as a function of temperature for the same levels. (c) The derivative
  $(d\sigma_{xy}/d\nu)^{max}$ as a function of temperature for the second
  electron and hole level.
  } 
  \label{Figure2}
\end{figure} 

The higher Landau levels in graphene have been shown to 
behave similar to the Landau level structure in traditional 
two-dimensional electron systems (2DEGs),\cite{Giesbers_PRL} 
and are therefore a good starting point for scaling measurements. The width $\Delta \nu$ 
and derivative $(d\sigma_{xy}/d\nu)^{max}$ as a function of temperature 
for the first and second electron and hole Landau level are shown in 
Figure~\ref{Figure2} at fixed magnetic fields between 5 and 30 T.

The widths $\Delta \nu$ and derivatives $(d\sigma_{xy}/d\nu)^{max}$ 
of these higher Landau levels show a clear dependence on temperature as can
be seen from Fig.~\ref{Figure2}. The widths $\Delta \nu$ of the first
electron and hole level at magnetic fields between 10 and 30 T show a 
power law behavior following $\Delta \nu \propto T^{\kappa}$. The scaling
exponents extracted from these data, for holes $\kappa=0.37\pm0.05$ and 
for electrons $\kappa=0.37\pm0.06$, are all identical within the error 
margins and show no evidence of a magnetic field dependence. The 
error is determined by the scattering of all the individual $\kappa$-values, 
the statistical error for each $\kappa$ is smaller. At 
$T<15$~K the curves in Fig.~\ref{Figure2}a flatten, 
which is an indication that the localization length becomes independent 
on temperature and becomes dominated by an intrinsic length 
scale,\cite{Koch2} possibly the 1~$\mu$m width of the sample.  

The linear behavior of the maximum derivatives 
$(d\sigma_{xy}/d\nu)^{max}$ in the Hall transition regions between the 
$\nu=\pm2$ and $\nu=\pm6$ plateaus are consistent with these observations, 
as shown in Fig.~\ref{Figure2}b. For both 
the first hole and the first electron level the curves show a power law
behavior with scaling exponents $\kappa=0.40\pm0.03$ and $\kappa=0.40\pm0.04$
respectively for all fields up to $B=25$~T. 

In the high field limit ($B=30$~T) the data shows a small reduction of the scaling 
exponent compared to the other fields.  
At these high fields the degeneracy of the Landau levels is partly lifted,\cite{Zhang} 
which leads to an additional minimum in $\sigma_{xx}$ and
a developing plateau in $\sigma_{xy}$ in the center of the Landau levels, here at
filling factor $\nu=\pm4$. Due to the broad Landau levels this splitting 
remains obscured in $\sigma_{xx}$ and $\sigma_{xy}$. However, their derivatives at the 
center Landau level positions are much more
sensitive to the onset of this splitting, which makes it 
difficult to extract reliable scaling data at the highest magnetic field.

At low temperatures, the $(d\sigma_{xy}/d\nu)^{max}$-curves measured between 
5 and 25 Tesla (see Fig.~\ref{Figure2}b) 
also flatten of, similar to what is observed in the width. This confirms that 
an intrinsic length apparently starts to dominate the localization length 
below $T=15$~K. 

Figure~\ref{Figure2}c shows $(d\sigma_{xy}/d\nu)^{max}$ for the second
hole and electron Landau levels. Despite of the limited temperature range
it does show a power law behavior with a scaling exponent of 
$\kappa=0.40\pm0.03$ and $\kappa=0.41\pm0.03$ for holes and electrons 
respectively. The width at the second Landau level is no longer
clearly distinguishable and therefore no data could be obtained on the 
scaling behavior of these levels from this technique.

When we assume the temperature exponent of the inelastic scattering length 
to be $p=2$, we obtain by the relation 
$\kappa=p/2\gamma$, a critical exponent $\gamma = 2.5 \pm 0.2$ for all the higher 
Landau levels combined. This value is in good agreement with the universal 
value $\gamma = 2.39\pm 0.1$ showing that localization of the higher Landau levels
in graphene follows a similar scaling behavior as traditional two-dimensional electron systems. 

\begin{figure}[tb]  
  \begin{center}
  \includegraphics[width=0.9\linewidth]{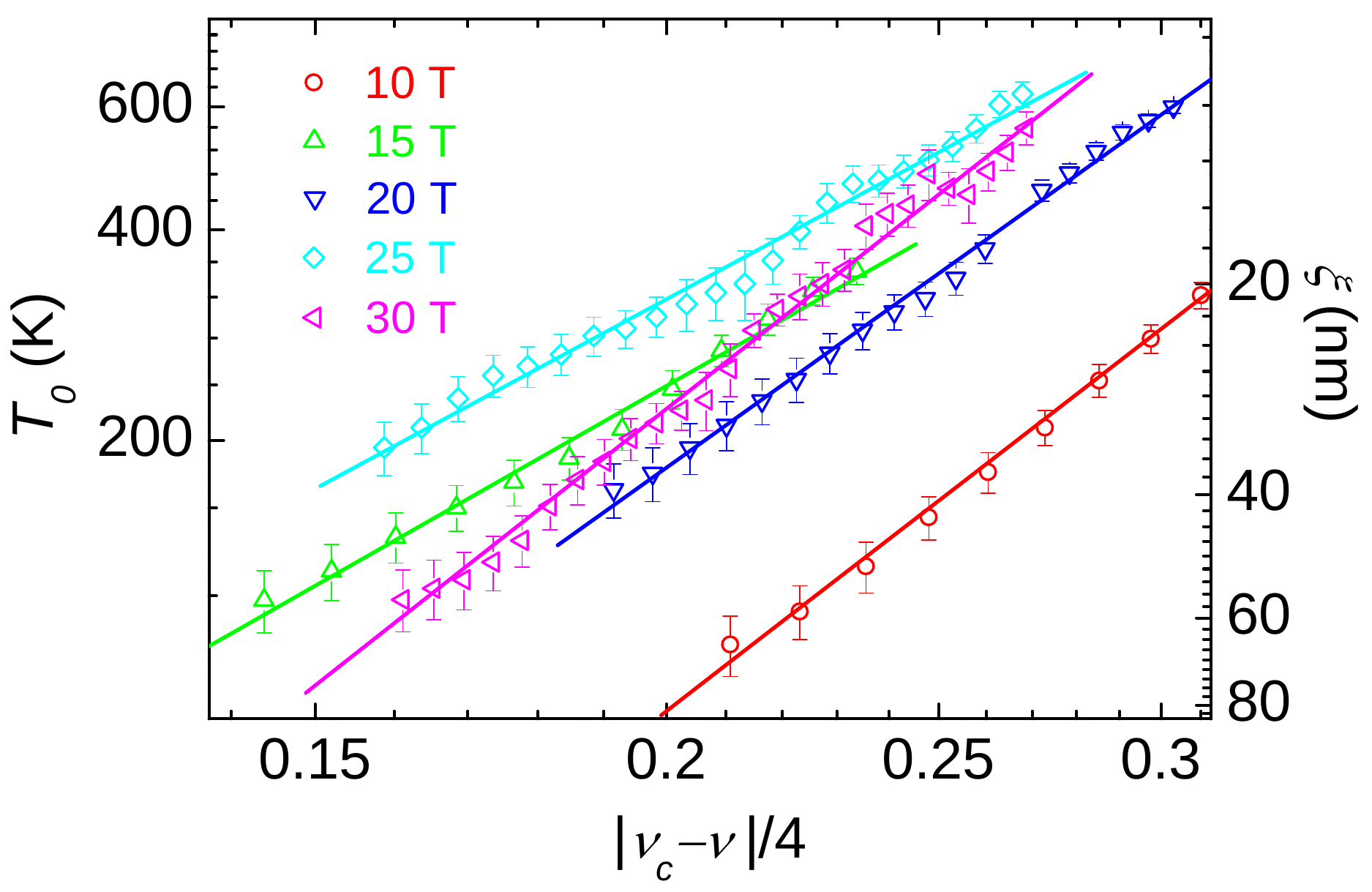}
  \end{center}
  \caption{(Color online) Scaling behavior in the tails of the 
  first electron Landau level for various magnetic fields as a function 
  of relative filling factor $|\nu_{c}-\nu|/4$ with $\nu=nh/eB$. The 
  factor $1/4$ in the $x$-axis accounts for the four-fold Landau level degeneracy. The 
  localization length on the right axis was calculated by 
  eq.~(\ref{chartemp}) with $C=1$.
  } 
  \label{Figure3}
\end{figure} 

These results are also consistent with the 
low temperature behavior of the conduction in the Landau level tails.\cite{Hohls} 
In these tails the conductivity decreases with decreasing 
temperature and disappears when the temperature is lowered to $T=0$
(see Fig.~\ref{Figure1}). When $kT$ is small enough to make the activation 
to the mobility edge and excitation across potential barriers to 
neighboring states improbable, conduction is governed by 
a variable range hopping type of conductivity. In this regime electrons 
or holes are able to tunnel between states within an energy range $kT$ 
leading to a slightly increased conductivity. The temperature dependence 
of the conductivity in this regime is 
given by\cite{Polyakov, Ebert, Briggs, Koch3, Furlan}
\begin{equation}\label{sigmatemp}
\sigma_{xx}=\sigma_{0}e^{-\sqrt{T_{0}/T}},
\end{equation}
with a temperature dependent prefactor $\sigma_{0}\propto 1/T$. The 
characteristic temperature $T_{0}$ is determined by the Coulomb 
energy and is inversely proportional to the localization length $\xi(\nu)$ 
at a particular filling factor $\nu$
\begin{equation}\label{chartemp}
T_{0}(\nu)= C\frac{e^{2}}{4\pi \epsilon \epsilon_{0} k_{B} \xi(\nu)},
\end{equation}
with $C$ a dimensionless constant in the order of unity and $\epsilon \approx 2.5$
is the effective dielectric constant for graphene on silicon dioxide.

Using equation (\ref{sigmatemp}) we can extract a value for the 
characteristic temperature $T_{0}$ from the temperature dependent 
conductivity measurements, which is directly related to the localization 
length via equation (\ref{chartemp}). Figure~\ref{Figure3} shows the 
results for the first electron Landau level as a function of the relative
filling factor $|\nu_{c}-\nu|$. 
A similar graph is found for the first hole level.
Combining these results with equation (\ref{loclength}) gives us a 
direct value for the critical exponent $\gamma = 2.6\pm 0.5$ for 
the first electron and hole Landau level. This is indeed in good 
agreement with the values obtained from the width $\Delta \nu$ and 
derivative $(d\sigma_{xy}/d\nu)^{max}$. 

It is interesting to note that the universality of the critical exponent
is preserved despite the four-fold degeneracy of the Landau levels in graphene
compared to lower degeneracies measured in other 2DESs. It 
was argued that degenerate levels would show a change in the critical 
exponent.\cite{hwang} The results presented here however 
show that
the Landau level degeneracy does not appear to be relevant.\cite{Zhao}   

\begin{figure}[tb]  
  \begin{center}
  \includegraphics[width=0.75\linewidth]{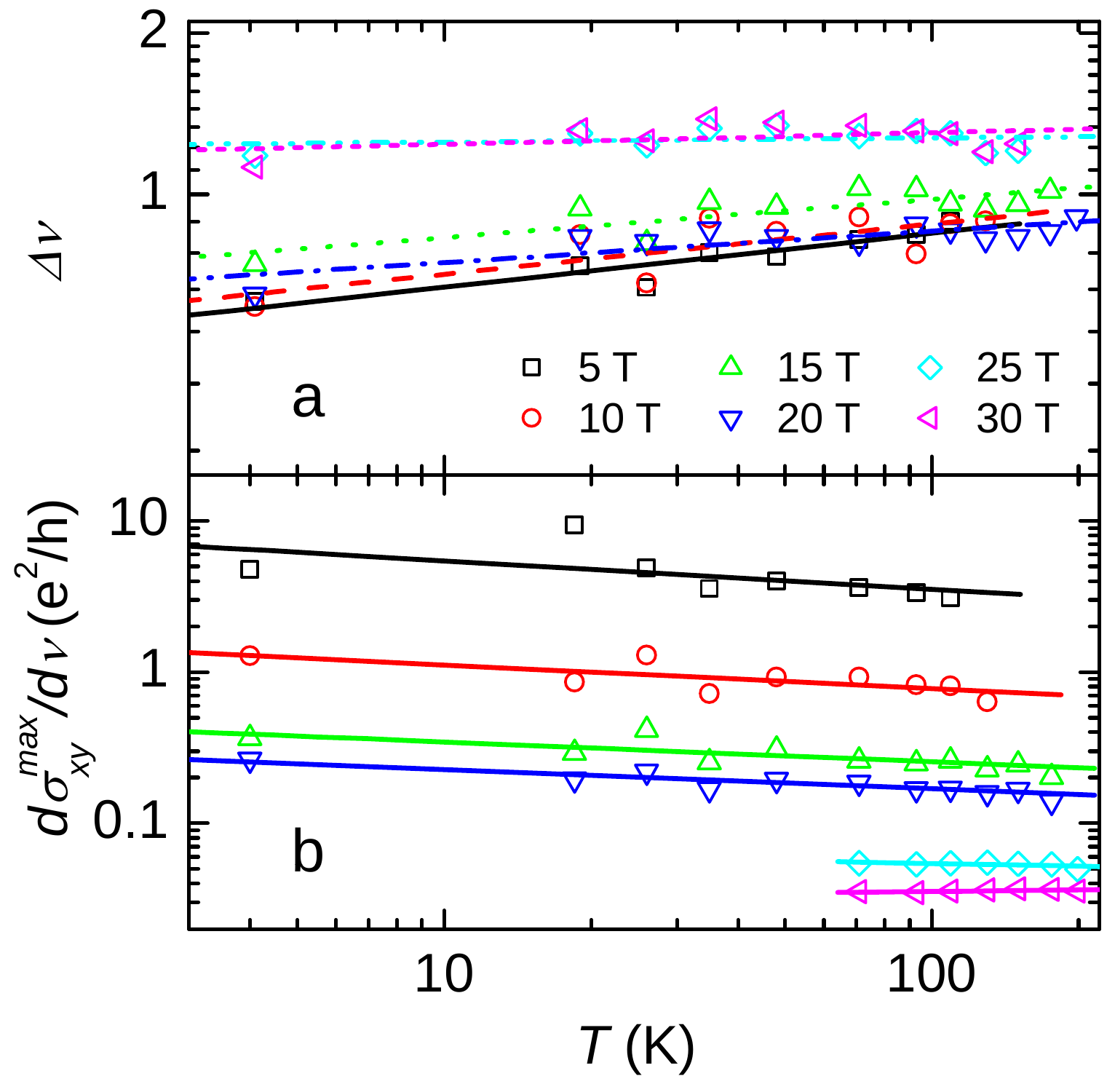}
  \end{center}
  \caption{(Color online) (a) $\Delta \nu$, distance between two extrema 
  in the derivative $d\sigma_{xx}/d\nu$, as a function of temperature for 
  the zeroth Landau level. 
  (b) The derivative $(d\sigma_{xy}/d\nu)^{max}$ as a function of temperature 
  for the zeroth Landau level. 
  } 
  \label{Figure4}
\end{figure} 

Let us now focus our attention to the zeroth 
Landau level. From a similar analysis as described before 
we obtain $\Delta \nu$ and 
$(d\sigma_{xy}/d\nu)^{max}$ as a function of temperature for magnetic 
fields between 5 and 30 T (see Fig.~\ref{Figure4}).  
Contrary to the higher Landau levels, at high fields neither the 
width (Fig.~\ref{Figure4}a) nor the derivative (Fig.~\ref{Figure4}b) 
of this zeroth level shows a clear dependence on temperature. 

Note that in the low temperature limit at high magnetic fields 
scaling measurements of the zeroth Landau level, especially in 
$(d\sigma_{xy}/d\nu)^{max}$, are obscured, similar to what we
observed for the $\nu=\pm4$ state in the first Landau level. In 
this regime the Landau level degeneracy is partly lifted and a 
$\nu=0$ state appears\cite{Abanin, Zhang, JosGap} and develops into a 
quantum Hall plateau at the CNP. Consequently $(d\sigma_{xy}/d\nu)^{max}$ 
will go to zero and is unrelated to any scaling behavior. 

The temperature independence of $\Delta \nu$ and 
$(d\sigma_{xy}/d\nu)^{max}$ in the zeroth Landau level 
suggest that the localization length is determined by an 
intrinsic length scale rather than a thermal length scale. 
The intrinsic length is in this case directly related to 
the localization length by 
$L_{int}= \xi = \xi_{0}|\Delta \nu/2|^{-\gamma}$ where the 
prefactor $\xi_{0}$ is proportional to the magnetic length 
$l_{B}=\sqrt{\hbar/eB}$ with a proportionality factor in the 
order of one.\cite{Huckestein} This provides an estimate for
the intrinsic length of the order of 10~nm.
This size is in reasonable 
agreement with the size of the intrinsic ripples in graphene, 
$L_{rip}\approx 10$~nm,\cite{Fasolino, Meyer_nat} suggesting 
that the origin of the temperature independent 
intrinsic length lies in ripple-induced 
localization.\cite{MishaSSC, MorozovPRL} Admittedly, 
numerous other scenarios, such as the existence of electron-hole 
puddles,\cite{ehpuddles} may also be considered as the
cause of a different localization mechanism in the zeroth Landau 
level compared to the higher levels.
(see e.g. also Ref.~\onlinecite{Ostrovsky} and references therein). 

To conclude we have shown that the width $\Delta \nu$ and the derivative
$(d\sigma_{xy}/d\nu)^{max}$ in the high Landau levels of graphene show a 
typical scaling behavior with an average $\kappa=0.40\pm0.02$ leading to 
a critical exponent $\gamma = 2.5 \pm 0.2$, where the temperature exponent 
was assumed to be $p=2$. These results are consistent with scaling
measurements in the VRH-regime yielding $\gamma = 2.6 \pm 0.5$.
In the zeroth Landau level no temperature dependence of the width nor of the 
derivative was observed in the measured range. We explain this result by a 
temperature independent intrinsic length leading to the absence of a
universal scaling behavior of the zeroth Landau level. 

Part of this work has been supported by EuroMagNET
under EU contract RII3-CT-2004-506239 and by the Stichting Fundamenteel
Onderzoek der Materie (FOM) with financial support from the
Nederlandse Organisatie voor Wetenschappelijk Onderzoek (NWO).
The work was also supported by Engineering and Physical Sciences Research
Council (UK), the Royal Society, the European Research Council (programs
"Ideas", call: ERC-2007-StG and "New and Emerging Science and Technology,"
project "Structural Information of Biological Molecules at Atomic
Resolution"), Office of Naval Research, and Air Force Office of Scientific
Research. The authors are grateful to Nacional de Grafite for supplying high
quality crystals of graphite.



\begin{thebibliography}{99}
\bibitem{note}
For low temperatures and high magnetic fields the resistance at the CNP starts to diverge which leads to
a splitting of the peak in $\sigma_{xx}$, see Ref.~\onlinecite{JosGap} and references therein. 
The data presented here still concernes the metallic regime where such a metal-insulator transition is not yet visible.
\end{thebibliography}
\end{document}